\documentstyle[prl,aps,preprint,psfig]{revtex}

\begin{document}
\tighten
\title{Cosmological Moduli Problem in Gauge-mediated 
Supersymmetry Breaking Theories}
\author{T. Asaka}
\address{Institute for Cosmic Ray Research, University of Tokyo,
  Tanashi 188, Japan}
\author{J. Hashiba}
\address{Department of Physics,  University of
  Tokyo, Tokyo 113, Japan}
\author{M. Kawasaki}
\address{Institute for Cosmic Ray Research, University of Tokyo,
  Tanashi 188, Japan}
\author{T. Yanagida}
\address{Department of Physics,  University of
  Tokyo, Tokyo 113, Japan}
\date{\today}

\maketitle

\begin{abstract}
A generic class of string theories predicts the existence of light
moduli fields, and they are expected to have masses $m_\phi$
comparable to the gravitino mass $m_{3/2}$ which is in a range of
$10^{-2}$keV--1GeV in gauge-mediated supersymmetry breaking
theories.  Such light fields with weak interactions suppressed by
the Planck scale can not avoid some stringent cosmological
constraints, that is, they suffer from `cosmological moduli
problems'.  We show that all the gravitino mass region $10^{-2}$keV
$\lesssim m_{3/2} \lesssim$ 1GeV is excluded by the constraints even if we
incorporate a late-time mini-inflation (thermal inflation).
However, a modification of the original thermal inflation model
enables the region $10^{-2}$keV $\lesssim m_{3/2} \lesssim$ 500keV to
survive the constraints. It is also stressed that the moduli can be
dark matter in our universe for the mass region
$10^{-2}$keV $\lesssim m_\phi \lesssim$ 100keV.

\end{abstract}



\section{Introduction}
%
Supergravity theories, which describe the low energy dynamics of 
string theories, generically have a large number of flat directions
in their field spaces \cite{Green}. We call the scalar fields corresponding to 
these flat directions moduli fields, or moduli, simply.
Moduli $\phi$ are expected to take their values of the order of
the Planck scale $M_{Pl}\simeq 1.2\times 10^{19}$GeV
in the very early universe, because it is the only scale appearing in
supergravity actions. At the later epoch in the universe's evolution
supersymmetry (SUSY) is spontaneously broken. SUSY- breaking effects
may lift the flat
potential for moduli through some non-perturbative dynamics and generate
their masses $m_{\phi}$ comparable to the gravitino mass :
$m_{\phi}\sim m_{3/2}$ \cite{Carlos-Casas-Quevedo-Roulet}. 

Moduli fields are generally accompanied by different kinds of cosmological
problems depending on the values of their masses. These problems are divided
into two classes discriminated by if or not the moduli decay takes place
in the early universe. That is, if it does the radiation produced
by the moduli decay may conflict with some cosmological observations, and
even if it does not a tremendous amount of moduli energy density itself
causes disaster. Let us make a crude estimation of moduli lifetime and
determine its dependence on $m_\phi$. From dimensional analysis the decay
width of the moduli is roughly
\begin{equation}
  \label{moduli-width}
  \Gamma_{\phi}\simeq N\frac{m_{\phi}^{3}}{M_{Pl}^{2}},
\end{equation}
where $N$ denotes the number of decay channels.
The lifetime is given by
\begin{equation}
  \label{moduli-lifetime}
  \tau_{\phi}=\Gamma_{\phi}^{-1}
    \simeq 1\times10^{20}\;{\rm sec}\;N^{-1}\left(\frac{m_{\phi}}
    {{\rm 10MeV}}\right)^{-3}.
\end{equation}

In conventional hidden sector models, the gravitino masses $m_{3/2}$
are about 1TeV, so from eq.(\ref{moduli-lifetime}) we see that
the moduli lifetime is much shorter than
the age of the present universe $\sim 3\times 10^{17}{\rm sec}$
for $m_{\phi}\sim m_{3/2}$. Therefore,
in these models one must worry about whether or not the radiation produced by
the moduli decay spoils the scenario of big bang nucleosynthesis.
The reheating temperature derived from the width (\ref{moduli-width}) is
\begin{equation}
  T_{R}\sim \sqrt{M_{Pl}\Gamma_{\phi}}\simeq
    0.3{\rm MeV}\;N^{\frac{1}{2}}\left(\frac{m_{\phi}}{10{\rm TeV}}\right)
    ^{\frac{3}{2}}.
\end{equation}
In order for the big bang nucleosynthesis to work, high enough reheating
temperature is needed ($T_{R}\gtrsim 10$MeV), and this requirement constrains
the moduli mass to be lower bounded : $m_\phi\gtrsim {\cal O}(10)$ TeV
\cite{Kawasaki-Moroi-Yanagida}\footnote{
It might appear that a large amount of entropy production by the moduli
decay predicts an extremely smaller value of the baryon-to-entropy ratio
than the one observed today even for $m_\phi \gtrsim$ 10 TeV,
but the Affleck-Dine mechanism for baryogenesis
\cite{Affleck-Dine} naturally explains the present value
\cite{Moroi-Yamaguchi-Yanagida}.
}.

However, this relatively large moduli mass
$m_\phi (\sim m_{3/2})\gtrsim 10$TeV
is realized only in a specific class of hidden sector models. Thus,
one is faced with a difficulty that the process of nucleosynthesis does not
proceed efficiently in a generic class of hidden sector models
\cite{Coughlan,Carlos-Casas-Quevedo-Roulet}. 
Lyth and Stewart proposed a mechanism so-called thermal inflation
\cite{Lyth-Stewart}
which solves this problem by diluting extensively the cosmic energy density
of the moduli and consequently decreasing the number of photons produced by
the moduli decay sufficiently so as not to upset the nucleosynthesis.

Gauge-mediated SUSY-breaking models \cite{Dine-Nelson-Shirman},
to which we pay our attention
especially in this paper, have the gravitino mass in a range of $10^{-2}$
keV-1GeV. As can be seen from eq.(\ref{moduli-lifetime}), in the mass range
$10^{-2}$keV$\lesssim m_\phi\lesssim$ 100 MeV
the moduli do not decay sufficiently
fast and are left in the present universe. Generically, moduli fields are
expected to take their vacuum-expectation values (vev's)
of the order $M_{Pl}$, which is so large that the moduli energy density
easily exceeds the critical density of the present universe. Although the
moduli lifetime (\ref{moduli-lifetime}) tells us that a class of these
models with 100 MeV $\lesssim m_\phi\lesssim$ 1 GeV
does not suffer from the moduli
problem of their too much amount of energy density, a constraint from the
cosmic $\gamma$-ray backgrounds is crucial in this mass region
\cite{K-Y,H-K-Y}.

Our main task in this paper is to apply the thermal inflation mechanism to
gauge-mediated SUSY-breaking models
\footnote{
This attempt has been first proposed in Ref.\cite{G-M-M}.
},
and argue the possibility that
these models could pass the above cosmological constraints.
We assume $m_\phi \simeq m_{3/2}$ throughout this paper.

Our analysis consists of two parts. In the first part (sections 2 and 3)
we adopt a thermal inflation model proposed by Lyth ane Stewart
\cite{Lyth-Stewart}. 
In this model there is a pseudo Nambu-Goldstone boson
(called R-axion) arising from
a spontaneous breakdown of an accidental R symmetry.
We show that the mass of R-axion is always much smaller than the flaton
mass in the gauge-mediated SUSY-breaking models.
Thus, the flaton decays mainly into
these R-axions and as a consequence the dilution of the moduli's energy
density becomes much milder. We show that there is no parameter region
surviving the cosmological constraints
in this type of thermal inflation model.

We modify the above thermal inflation model, in the second part (section 4),
by adding a small explicit breaking term of the R symmetry to suppress
the R-axion decay of the flaton. We find that a parameter region for
the moduli mass, $10^{-2}$keV$\lesssim m_\phi\lesssim 500$keV
\cite{H-K-Y}, survives
the cosmological constraints. It should be noted here that
the Affleck-Dine baryogenesis does work for these gravitino mass region
\cite{G-M-M}.
We also stress that the moduli themselves could be the dark matter in the
present universe for the region,
$10^{-2}$keV$\lesssim m_\phi\lesssim$ 100 keV.
This may be a very crucial observation, 
since there is no dark matter candidate
beside the moduli themselves because of the substantial dilution of
any relic particle abundance by
the late-time thermal inflation
\footnote{
Peccei-Quinn axion with high values of the decay constant
$F_a \simeq 10^{15}$--$10^{16}$ GeV may be another candidate
for the dark matter\cite{K-Y-2}. See also Ref.
\cite{L-S-S-S}.
}. 
The last section is devoted to discussion and conclusions.

%
\section{Thermal inflation}
In this section we review a thermal inflation model which was
proposed by Lyth and Stewart\cite{Lyth-Stewart}.
Inevitable ingredients of the model are flaton fields,
which are characterized by their 
`almost flat' potentials and `large' vacuum-expectation values 
(vev's).
Here, `almost flat' means that the roots of the second rank derivatives of
potentials at their minima, namely the masses of corresponding
flaton particles, are of the order of SUSY-breaking scale
and the flaton fields have `large' vev's which are much larger than 
the scale.
For simplicity,
let us concentrate on the case in which only one flaton field exists, even
though more than one flaton cases are also valuable to analyze.

In order to guarantee the flatness of potential, 
we assume that the flaton's potential possesses some exact
(or at least approximate) global symmetries. 
If not, nonrenormalizable higher order interactions 
would induce a tadpole term through the SUSY-breaking effects.
One familiar example of
such global symmetries is $U(1)_{R}$ symmetry, 
but it should be explicitly broken to some discrete symmetry, say
${\bf Z}_{n}$, by a constant term $C$ in the superpotential 
which is required to cancel the cosmological constant. 
Thus, in this paper we postulate that
the superpotential for the flaton exhibits ${\bf Z}_{n+3}\;(n\geq 1)$
symmetry and takes a form
\footnote{
Even if one assumes an approximate $U(1)_{R}$ symmetry instead of
${\bf Z}_{n+3}$, one reaches the same conclusion as in this paper,
since the higher power terms $(k=2,\cdots,\infty)$ in the superpotential
(\ref{superpotential}) are practically negligible as seen below.
}
\begin{equation}
  \label{superpotential}
  W ~ = ~ C ~ + ~ 
  \sum_{k=1}^{\infty}\frac{\lambda_{k}}{(n+3)k}\frac{X^{(n+3)k}}
  {M_{*}^{(n+3)k-3}},
\end{equation}
where $X$ is the flaton chiral superfield 
and $\lambda_{k}\;(\lambda_{1}=1)$ coupling constants.
$M_{*}$ denotes the cut-off scale of the models we consider%
\footnote{
In the original paper\cite{Lyth-Stewart} the cut-off scale $M_\ast$ 
is taken to be at the Planck scale.
Here, we regard it as a free parameter to make a general analysis.
}.
As will be verified later the vev of the superpotential 
(\ref{superpotential}) is
dominated by the constant term $C$, and hence we take
\begin{equation}
  \label{eq-c}
  |C| \simeq M_{G}^{2}m_{3/2},
\end{equation}
to cancel the vacuum energy. Here, $M_G$ is the reduced Planck scale,
$M_G=M_{Pl}/\sqrt{8\pi}\simeq 2.4\times 10^{18}$GeV.

Then, the effective potential of the flaton $X$ at low energy scale
is represented as
\begin{equation}
  \label{flaton-potential}
  V_{eff}(X) ~\simeq~ 
  V_{0}~ - ~ m_{0}^{2}|X|^{2} 
  ~ + ~ \frac{n}{n+3}\frac{ C }{M_G^2 M_{*}^{n}}
    (X^{n+3}+X^{*n+3})
  ~ + ~ \frac{1}{M_{*}^{2n}}|X|^{2n+4},
\end{equation}
where we have used the same letter $X$ for the flaton complex scalar field
as for the corresponding superfield, hoping that readers should not be
confused.
Note that the dynamics of the flaton field $X$
is governed by the leading term in the superpotential 
(\ref{superpotential}) for $X\ll M_*$
and we have neglected higher order terms in eq.(\ref{flaton-potential})
since the vev of the flaton is much smaller than the cut-off scale $M_\ast$.
$V_{0}$ is determined by the requirement that the cosmological
constant vanishes
\footnote{$V_{0}$ is at the tree-level given by
\begin{equation}
  V_{0}\simeq |F|^{2}-\frac{3}{M_G^2}|C|^{2},
\end{equation}
where $F$ is the SUSY-breaking $F$ term.
}. 
The quadratic term in eq.(\ref{flaton-potential}) 
is induced by SUSY-breaking effects \cite{M-S-Y}
and we assume the mass squared at the origin $X\simeq 0$
to be negative and of the order of SUSY-breaking scale.
The vev of the flaton is given by
\begin{equation}
  \label{flaton-vev}
  \langle X\rangle ~\equiv~ M
  ~ \simeq ~ \left(\frac{1}{n+2}\right)^{\frac{1}{2(n+1)}}
    (m_{0}M_{*}^{n})^{\frac{1}{n+1}},
\end{equation}
where we have neglected ${\cal O}(m_{3/2}/m_0)$ terms.
(Remember that the formulae given below are also valid only up to
${\cal O}(m_{3/2}/m_0)$.
See the paragraph following eq.(\ref{axiondecay}).)
Then $V_{0}$ is given by
\begin{equation}
  V_{0} ~\simeq~ \frac{n+1}{n+2}m_{0}^{2}M^{2}.
\end{equation}
Here, reflecting the ${\bf Z}_{n+3}$ symmetry 
the degenerate minima cause a potential domain wall problem.
We will come back this point in section 4.

The superpotential (\ref{superpotential}) has 
an approximate $U(1)_R$ symmetry if we neglect the higher order terms
and then the imaginary part of $X$
becomes a pseudo-Goldstone boson called R-axion. 
The R-axion receives a mass from the constant term $C$ which breaks
the $U(1)_R$ symmetry explicitly \cite{Bagger}.
If we parameterize $X$ as
\begin{equation}
  X=\left(\frac{1}{\sqrt{2}}\chi+M\right)\exp\left(\frac{ia}{\sqrt{2}M}
          \right),
\end{equation}
with $\chi$ representing the real flaton field and $a$ the R-axion,
then the flaton and the R-axion mass squared
are estimated as%
\footnote{
Here we choose the phase of the constant term in eq.(\ref{eq-c})
as $C \simeq - M_G^2 m_{3/2}$ so that the R-axion mass squared 
is positive.
}%
\begin{eqnarray}
  m_{\chi}^{2} &\simeq& 2(n+1)m_{0}^{2}, \label{flaton-mass}\\
  m_{a}^{2} &\simeq& \frac{n(n+3)}{\sqrt{n+2}}m_{0}m_{3/2}
  ~\simeq~ \frac{ n (n+3) }{ \sqrt{ 2 (n+1) (n+2) } } m_\chi m_{3/2}. 
  \label{axion-mass}
\end{eqnarray}

The thermal inflation occurs if the flaton field does not sit at the
true minimum of the potential but at the origin 
in the early universe. To realize this initial condition for the thermal
inflation the flaton must have 
interactions with other fields in the thermal bath of the universe%
\footnote{
For example, a Yukawa interaction as 
$W = g X \xi \overline{\xi}$ is sufficient.
The fields $\xi$ and $\bar{\xi}$ receive a mass $m_\xi \simeq g M$ when the 
flaton sits at the true minimum, but for the flaton field values near
the origin the fields $\xi$ and $\bar{\xi}$ become light and
could be in the thermal bath
if they couple to the particles in the standard model and the temperature
$T$ is larger than the mass $m_\xi$.
}.
From the finite temperature effects, the effective potential in the 
early universe takes a form as
\begin{equation}
  \label{effective-potential}
  V_{eff}(X) ~\simeq~ V_{0}~ +~ (cT^{2}-m_{0}^{2})|X|^{2}
   ~-~ \frac{n}{n+3}
          \frac{m_{3/2}}{M_{*}^{n}}(X^{n+3}+X^{*n+3})
          ~+~ \frac{1}{M_{*}^{2n}}|X|^{2n+4},
\end{equation}
where $c$ is a constant of ${\cal O}(1)$ and $T$ is the temperature
of the universe.
Then, at high temperature $T \gtrsim T_c \sim m_0$ 
the flaton field sits near the origin and produces the vacuum energy $V_{0}$.

If the energy of the universe is dominated by the radiation
just before the thermal inflation, 
the vacuum energy $V_0$ becomes comparable to the radiation energy 
at the cosmic temperature $T_\ast \sim V_0^{1/4}$.
Thus, for $T_c \lesssim T \lesssim T_\ast$,
the vacuum energy of the flaton field dominates the energy of the 
universe and a mini-inflation, i.e.
thermal inflation takes place\cite{Lyth-Stewart}.
On the other hand, 
if the moduli oscillations dominate the energy of the universe before
the thermal inflation,
the temperature at the beginning of thermal inflation
is estimated as $T_\ast \sim ( V_0^2 /(m_\phi M_G) )^{1/6}$.
Here $m_\phi$ are the moduli masses.

In considering the history of the universe after the thermal inflation,
the flaton decay is crucial since it is most responsible for the 
entropy production.
Here we shall list the relevant decay channels and compute
quantitatively the decay rate for each channel which is needed to trace
the physics following the thermal inflation epoch.

The only possible renormalizable interaction of the flaton with SUSY
standard model particles is
\begin{equation}
  \label{w-higgs}
  W=\lambda X H\bar{H},
\end{equation}
where $H$ and $\bar{H}$ are Higgs chiral supermultiplets. When the flaton field
develops the vev, $\langle X\rangle =M$, the Higgs multiplets acquire a mass
$\lambda M$, and for this to be at most the electroweak scale the coupling
constant $\lambda$ has to be set so small, $\lambda \lesssim \mu_{H}/M$. 
Here $\mu_{H}$ is the SUSY-invariant mass for the Higgs multiplets.
If the flaton mass is large enough,
the flaton decays into a pair of Higgs fields%
\footnote{
Here, Higgs field denotes a Higgs boson or a Higgsino.
}.
The decay width is represented as
\begin{equation}
  \Gamma_{\chi\rightarrow 2h} ~\simeq ~
  C_h \frac{ 1 }{ 16 \pi } \frac{ m_\chi^3 }{ M^2 },
  \label{chi-2higgs}
\end{equation}
where $C_h$ is a constant parameter satisfying $C_h \lesssim {\cal O}(1)$%
\footnote{
For example, the coupling $C_h$ is $C_h = ( \lambda M/ m_\chi )^4$ 
in the case the flaton decays into two Higgs bosons.
And $C_h = ( \lambda M / m_\chi )^2$
for the case the flaton decays into two Higgsinos.
Thus in this analysis we shall assume that 
the coupling $C_h$ is a free parameter of  $C_h \lesssim {\cal O}(1)$,
since $\lambda \lesssim \mu_{H}/M$.
},
and we have ignored the masses of the Higgs fields.

Even if the above decay process would be kinematically forbidden,
the flaton could decay into two photons through one-loop diagrams\cite{H-K-Y}%
\footnote{
This decay process is induced by the Higgs fields loop diagram 
through the interaction (\ref{w-higgs})
or by charged $\xi$ loop through 
the Yukawa interaction which is required 
in order that the flaton sits around the origin 
during the thermal inflation.
If $\xi$ has a color charge, 
the decay into two gluons occurs and becomes a dominant decay
for the flaton mass $m_{\chi}\geq$ a few GeV.
}.
The width of this decay channel is given by
\begin{eqnarray}
  \Gamma_{\chi\rightarrow 2\gamma} 
   &\simeq& \frac{1}{8\pi}\left(\frac{\alpha_{em}}
           {4\pi}\right)^{2}\frac{m_{\chi}^{3}}{M^{2}}.
        \label{chi-2photon}
\end{eqnarray}
Through the above decay processes, 
the flaton energy is transferred to the radiation
and reheats the universe
\footnote{
A possible decay into two pairs of bottom and antibottom quarks is
strongly suppressed by phase volume effects.
}.

One should note that 
the flaton can decays into two R-axions,
if kinematically allowed.
The decay rate is calculated as
\begin{eqnarray}
  \Gamma_{\chi\rightarrow 2a} \simeq \frac{1}{64\pi}\frac{m_{\chi}^{3}}{M^{2}}.
    \label{chi-2axion}
\end{eqnarray}
Here, we have neglected the R-axion mass.
The R-axions produced by this process successively decay into two photons
similarly to the flaton decay 
and its decay rate is estimated as
\begin{equation}
  \Gamma_{a\rightarrow 2\gamma}
   ~\simeq~ \frac{1}{8\pi}\left(\frac{\alpha_{em}}{4\pi}
    \right)^{2}\frac{m_{a}^{3}}{M^{2}}.
  \label{axiondecay}
\end{equation}

In gauge-mediated SUSY breaking theories 
the predicted gravitino mass range ($m_{3/2}\simeq 10^{-2}$keV--1GeV)
indicates that the flaton decay into two R-axions
is most likely allowed, since the flaton may obtain a mass of the order
of the SUSY-breaking scale ($m_{\chi}\simeq$ 10GeV--1TeV)
and the R-axion has a mass $m_a\sim \sqrt{m_{3/2}m_{\chi}}$
(see eqs.(\ref{flaton-mass}) and (\ref{axion-mass}))
\footnote{
\label{hidden-sector}
In Ref.\cite{Lyth-Stewart} hidden sector models of the SUSY breaking
are considered where the gravitino has a mass
$m_{3/2} \simeq m_\phi \simeq$ 100 GeV--1 TeV.
In this case the R-axion has a mass of the order
$m_\phi$ (see eq.(\ref{axion-mass})) and the flaton decay into R-axions
may not be allowed.
}. 
Thus, in the following, we consider the case
that the flaton can decay into standard model particles
(Higgs fields or photons) and two R-axions.
(For the case that the R-axion decay is forbidden,
see section 4.)

Now we are ready to describe the history after the thermal inflation.
Our aim here is to estimate the entropy 
produced by the flaton decay processes.
At the temperature $T \sim T_c$ the flaton field starts to roll
down to the true minimum of the potential (\ref{flaton-potential})
and then oscillates around it.
When the Hubble parameter becomes of the order of 
total width of the flaton,
the flaton $\chi$ decays into both standard model particles
and R-axions.
The energy of R-axions can not be transferred to the radiation
at this time because it has only weak interaction 
suppressed by $1/M$
with particles in the thermal bath.
Then only the energy of the standard model particles is transfered to
the radiation to reheat the temperature of the universe $T_{R,SM}$
at the flaton decay epoch.
The ratio of the entropy densities just before to after the flaton decay 
is estimated as
\begin{eqnarray}
  \label{delta-sm}
  \Delta_{SM} ~\simeq~ 1+( 1 - \epsilon_a )\frac{ 4 }{ 3 }
  \frac{ V_0 }{ ( 2 \pi^2/45 ) g_\ast T_c^3 T_{R,SM} },
\end{eqnarray}
where $g_\ast$ is the effective number of degrees of freedom
and $\epsilon_a$ denotes the branching ratio of
the flaton decay into two R-axions.

Just after the flaton decay,
the energy density of the R-axion per the entropy density
is $\epsilon_a\left(\frac{ V_0 }{ ( 2 \pi^2/45 ) g_\ast T_c^3 }\right)
\frac{1}{\Delta_{SM}}$.
The energy densities of the R-axion and radiation
are both diluted by the expansion at the same rate $R^{-4}$,
where $R$ is the scale factor of the universe.
But after the R-axions become non-relativistic particles,
their energy density $\rho_a$ is diluted at $R^{-3}$ and dominates over
the energy of the radiation unless $\epsilon_a \simeq 0$.

The R-axion decay into two photons occurs at the Hubble parameter
$H \sim \Gamma_{a \rightarrow 2\gamma}$ 
and the universe is reheated again with the temperature $T_R$.
At this time, the R-axion decay increases the entropy 
by a factor
\begin{eqnarray}
  \label{delta-a}
  \Delta_{a} ~\simeq~ 1+\epsilon_a\frac{ 4 }{ 3 }
  \frac{ V_0 }{ ( 2 \pi^2/45 ) g_\ast T_c^3 T_R }\frac{1}{\Delta_{SM}}
  \left( \frac{ 2 m_a }{ m_\chi } \right).
\end{eqnarray}

Then, all the vacuum energy of the thermal inflaton (i.e. flaton)
is eventually transferred to the radiation. 
The reheating temperature $T_{R}$ is estimated from 
eq.(\ref{axiondecay}) as
\begin{eqnarray}
  T_R &\simeq& 1.1 \times 10^{-4} ~\frac{m_a^{3/2}M_G^{1/2}}{M}.
  \label{reheatingtemp}
\end{eqnarray}
And the thermal inflation increases the 
entropy of the universe by a factor
\begin{eqnarray}
  \label{delta-s}
  \Delta = \Delta_{SM}\cdot\Delta_{a}~\simeq~
  1+(1-\epsilon_a)\frac{ 4 }{ 3 }
  \frac{ V_0 }{ ( 2 \pi^2/45 ) g_\ast T_c^3 T_{R,SM}}
  +\epsilon_a\frac{ 4 }{ 3 }\frac{ V_0 }{ ( 2 \pi^2/45 ) g_\ast T_c^3 T_{R}}
  \left( \frac{ 2 m_a }{ m_\chi } \right).
\end{eqnarray}

As we have described the thermal inflation reproduces the entropy
at the late-time of the universe's evolution
which dilutes substantially the cosmic energy densities of
any relic particles such as moduli fields. 
The dilution factor is given by eq.(\ref{delta-s})
when the two R-axion decay of the flaton is allowed. The dilution factor
in the other case will be given in section 4.
We should note that the requirement $T_R \gtrsim$10 MeV in
eq.(\ref{reheatingtemp}) leads to the upper bound for $M$
which justifies our ansatz in eq.(\ref{eq-c}).

%
\section{Moduli problem in gauge-mediated SUSY breaking theories}
In this section, we consider the cosmological moduli problem in
gauge-mediated SUSY breaking theories.
The predicted gravitino mass 
$m_{3/2} \simeq 10^{-2}$ keV--1 GeV
indicates that the string moduli fields 
may have lifetimes longer than the age of the universe.
Then one clear problem arises that the energy densities of the moduli 
overclose the universe.
Moreover, it has been pointed out recently that 
the contributions to the cosmic $X(\gamma)$-ray background
from the moduli decays are very dangerous\cite{K-Y,H-K-Y}.
The only known possibility to solve these problems 
is the thermal inflation.
Therefore, we examine whether the thermal inflation discussed in
the previous section can solve these moduli problems.

\subsection{The energy density of string moduli}
First, we discuss the problem of the 
cosmic energy density of the moduli fields.
To derive conservative cosmological constraints,
we assume that at least one modulus field exists with
a mass $m_\phi \simeq m_{3/2}$ and it has 
an initial value of the order of the gravity scale $M_G$.
(Here, we have chosen vev of the modulus field at the origin).
Generalization to the case that many moduli fields exist is straightforward.

Let us show by explicit calculation that the modulus energy density
exceeds substantially the critical density of the present universe
if the modulus is stable.
When the expansion rate of the universe becomes of the order of the 
the modulus mass, $H \sim m_\phi$,
the modulus field $\phi$ starts to oscillate
around the minimum of the potential
with the initial amplitude $\phi_0$ of the order $M_G$.
At that time the energy density of the modulus coherent oscillation is 
\begin{eqnarray}
  \rho_\phi ~=~ \frac{1}{2} m_\phi^2 \phi^2_0,
\end{eqnarray}
and the energy density of radiation is of the same order of $\rho_\phi$.
Thus, the temperature of the universe when the modulus starts
to oscillate is estimated as 
\begin{eqnarray}
  T_\phi ~&\simeq&~ \left( \frac{90}{\pi^2 g_{\ast}} \right)^{1/4}
               \sqrt{ M_G m_\phi },
\\
     &\simeq&~ 7.2 \times 10^{8} ~\mbox{GeV}
     \left( \frac{m_\phi}{1 \mbox{GeV}} \right)^{1/2}.
     \label{tphi}
\end{eqnarray}
Here, we have assumed that the modulus coherent oscillation begins
after reheating of the ordinary inflation completes,
and that its reheating temperature is higher than $T_\phi$%
\footnote{
In some models of inflation, such as a chaotic or hybrid inflation,
one may easily have the reheating temperature higher than  $T_\phi$
in eq.(\ref{tphi})\cite{Linde}.
}.
Then the energy density per the entropy density is given by 
\begin{eqnarray}
  \label{rhophi-bb}
  \frac{\rho_\phi}{ s }
    &\simeq&
  \frac{ m_\phi^2 \phi_0^2 /2}{ (2\pi^2/45)g_{\ast} T_\phi^3 },\\
    &\simeq&
    0.9 \times 10^8 ~ \mbox{GeV} 
      \left( \frac{ m_\phi }{ 1~\mbox{GeV} } \right)^{1/2}
      \left( \frac{ \phi_0 }{ M_G } \right)^2.
      \label{rhophi-bb2}
\end{eqnarray}
This ratio takes a constant value until the present if no 
entropy is reproduced, 
because the densities of the energy and of
the entropy are diluted at the same rate as $R^{-3}$
as the scale factor $R$ increases.

On the other hand, the critical density of the present universe
is given by
\begin{eqnarray}
    \label{critical-density}
    \frac{ \rho_c}{ s } ~=~
    3.6 \times 10^{-9} ~ h^2 ~ \mbox{GeV},
\end{eqnarray}
where $h$ is the present Hubble parameter in units of
100 km/sec/Mpc.
When we consider gauge mediated SUSY-breaking theories
where the predicted modulus mass is $m_\phi \simeq m_{3/2}$
$\simeq$ $10^{-2}$ keV-1 GeV,
one can see from eq.(\ref{rhophi-bb2}) that
the energy density of the modulus coherent oscillation
overcloses the universe if the modulus is stable until now.
The thermal inflation increases the entropy of the universe 
by the factor $\Delta$ as shown in eq.(\ref{delta-s}) 
and dilutes the energy density of the modulus given by eq.(\ref{rhophi-bb}).

Now we adopt the thermal inflation model in the previous section and
estimate the minimum value of the present 
energy density of the modulus field $\phi$.
The relevant dynamics is determined by 
two mass scales $m_0$, $M_\ast$ and the branching ratio $\epsilon_a$ for a
given gravitino mass $m_{3/2}$ ($\simeq m_\phi$).
Since $\Delta$ in eq.(\ref{delta-s}) takes its maximum at
$\epsilon_a = 1$ as far as $m_a \ll m_\chi$,
we put $\epsilon_a = 1$
to obtain the most efficient dilution factor
\footnote{
When the flaton decay into Higgs fields is forbidden,
we have $\epsilon_a \simeq 1$. When it is allowd, $\epsilon_a$
depends $C_h$ in eq.(\ref{chi-2higgs}). However, we find that
the dilution factor $\Delta$ takes the maximum value at $C_h \simeq 0$
which means $\epsilon_a \simeq 1$.
}.
In the following analysis we take two free parameters
$m_\chi$ and $T_R$ instead of $m_0$ and $M_\ast$, and 
search the minimum energy density of the modulus in the present universe.

The amount of the present energy density of modulus takes different forms
depending on whether the modulus field begins to oscillate before
the thermal inflation or after the end of that. 
For the moment we assume that the modulus coherent oscillation starts
before the thermal inflation, i.e. $m_{\phi}>H_{TI}$.
Here, $H_{TI} \simeq \sqrt{V_0}/(\sqrt{3} M_G)$ is the Hubble parameter
during the thermal inflation.
Then, from eq.(\ref{rhophi-bb}) the present energy density
of such a modulus (``big-bang'' modulus) is 
\begin{eqnarray}
  \label{pre-rhobb}
  \left( \frac{\rho_\phi}{ s } \right)_{BB} 
    ~\simeq~
  \frac{ m_\phi^2 \phi_0^2 /2}{ (2\pi^2/45)g_{\ast} T_\phi^3 }
  ~ \frac{ 1 }{ \Delta }.
\end{eqnarray}
Furthermore, the modulus energy 
is reproduced after the thermal inflation.
Because, during the thermal inflation, 
the modulus sits at the minimum of the potential 
which is shifted from its true vacuum by an amount of
$\delta \phi \sim (V_0/m_\phi^2 M_G^2) \phi_0$ \cite{Lyth-Stewart} and 
restarts to oscillate around the true minimum with an amplitude
$\delta \phi$ after the end of the thermal inflation.
Then the present energy density of this ``thermal-inflation'' modulus is 
estimated as
\begin{eqnarray}
  \label{pre-rhoti}
  \left( \frac{\rho_\phi}{ s } \right)_{TI} 
    ~\simeq~
    \frac{ m_\phi^2 (\delta\phi)^2 /2 }
         {(2 \pi^2 /45 ) g_\ast T_c^3 }
 ~ \frac{ 1 }{ \Delta }.
\end{eqnarray}
Therefore, the present total energy density of the modulus
with mass $m_\phi > H_{T.I.}$ is\cite{H-K-Y}
\begin{eqnarray}
  \label{ab-1}
  \left(\frac{ \rho_\phi }{ s }\right)_0 ~
  &\simeq& ~
  \mbox{ max } 
  \left[ ~
   \left( \frac{\rho_\phi}{ s } \right)_{BB}, ~
   \left( \frac{\rho_\phi}{ s } \right)_{TI} ~
  \right]
   ~ \gtrsim ~ \left( \frac{\rho_\phi}{ s } \right)_{BB}, \nonumber\\
  &\simeq& ~
    6.1 \times 10^8 ~
  \frac{ (n+1)(n+2)^2 }{ n^2 (n+3)^2 }
  \left( \frac{ T_c }{ m_\chi } \right)^3
  \left( \frac{ \phi_0 }{ M_G } \right)^2
  \frac{ T_R^3 }{ M_G^{1/2} m_{3/2}^{3/2} } \label{pre-rhobb2}
\end{eqnarray}
Here, notice that it depends only on $T_R$ and not on $m_\chi$
since $T_c \sim m_\chi$.
The reheating temperature should satisfy
$T_R \gtrsim 10$ MeV to maintain the success of the
big bang neucleosynthesis.
Then if we choose the minimum value of $T_{R} \simeq$ 10 MeV
in eq.(\ref{pre-rhobb2}), 
we obtain the lowest value of the present modulus energy density
for $T_c \simeq m_\chi$ and $\phi_0 \simeq M_G$,
\begin{equation}
  \left(\frac{ \rho_\phi }{ s }\right)_0 ~\gtrsim ~
  4.0\times 10^{-7} ~\mbox{GeV}~
  \frac{ (n+1)(n+2)^2 }{ n^2 (n+3)^2 }
  \left( \frac{ m_{3/2} }{ 1~\mbox{GeV} }\right)^{-3/2}.
\end{equation}

On the other hand, 
if the modulus mass is smaller than $H_{TI}$,
then the modulus oscillation begins after the end of the thermal inflation
with the amplitude $\phi_0 \sim M_G$.
The entropy of the universe when the modulus starts to oscillate
is estimated as
\begin{eqnarray}
    s ~\simeq~ \frac{ 2 \pi^2}{ 45 }g_{\ast} T_c^3 
    ~\frac{ 3 M_G^2 m_\phi^2 }{ V_0 },
\end{eqnarray}
where we have used the fact that the entropy is diluted by 
the expansion of the universe at the rate $R^{-3}$.
Then the present abundance of the modulus is%
\footnote{
Here the modulus oscillation is assumed to start at least 
before the R-axion decays, i.e.,
$m_\phi \gtrsim \Gamma_{a \rightarrow 2\gamma } 
~\simeq ~1.1 ~T_R^2/M_G$
$\simeq 4.5 \times 10^{-23}$ GeV ($T_R$/10 MeV$)^2$.
\label{assumption}
}
\begin{eqnarray}
  \label{ab-2}
  \left( \frac{\rho_\phi}{ s } \right)_{0} 
   &\simeq &
  \left( \frac{ 15 }{4 \pi^2 g_{\ast} } \right)
  \frac{ \phi_0^2 V_0 }{ M_G^2 T_c^3 } 
~ \frac{ 1 }{ \Delta }, \\ 
   &\simeq &
   \frac{1}{16}
   \frac{ \left[ 2 (n+1)(n+2) \right]^{1/4} }{ \sqrt{ n (n+3) } }
   \left( \frac{ \phi_0 }{ M_G } \right)^2
   \frac{ m_\chi^{1/2} T_R }{ m_{3/2}^{1/2} }.
\end{eqnarray}
In this case, the mass of flaton $\chi$, (since 
$m_\phi < H_{TI}$) should satisfy
\begin{eqnarray}
  m_{\chi} &\gtrsim& 3.5  \times 10^2 ~
  \frac{ (n+2)^{1/2} (n+1)^{3/14}  }
       { [ n(n+3) ]^{3/7} } ~
  M_{G}^{2/7} ~ m_{3/2}^{1/7} ~
  T_{R}^{4/7},
\end{eqnarray}
and this gives the lower bound as
\begin{eqnarray}
  \label{min-rho-ati}
   \left( \frac{ \rho_\phi }{ s } \right)_{0}
   &\gtrsim&
    1.6 ~\mbox{GeV}~
    \frac{ (n+2)^{1/2} (n+1)^{5/14} }{ [n(n+3)]^{5/7} }
    \left(\frac{ 1 ~\mbox{GeV} }{ m_{3/2} }\right)^{3/7},
\end{eqnarray}
for the minimum value of the reheating temperature 
$T_R \simeq $ 10 MeV and $\phi_0 \simeq M_G$.
Here we have neglected the region where the gravitino mass is less than 
\begin{eqnarray}
    \label{m32-cr2}
    m_{3/2} \simeq 2.4 \times 10^{6} ~
    \frac{ 1 }{ (n+1)^{1/10} [n(n+3)]^{6/5} }
    \frac{ T_R^{8/5} }{ M_G^{3/5} },
\end{eqnarray}
since the vev of the flaton $M$ exceeds the cutoff scale $M_\ast$
and our effective treatment of the flaton potential 
breaks down.

To compare our result with the critical density of the present universe,
we show the obtained lower limit for
\begin{eqnarray}
        \Omega_{\phi} h^2 \equiv \frac{\rho_{\phi} h^2}{ \rho_{c} }.
\end{eqnarray}
in Fig.1. 
In this analysis we have assumed that the modulus is stable and hence
this figure shows the cosmic energy density of the stable modulus'
coherent oscillation. We find that it exceeds largely
the critical density of the present universe.
Notice that in this figure we take the case of $n=1$.%

So far, we have assumed that the modulus field $\phi$ is stable.
However, it is not valid, since the modulus field $\phi$ may couple to
the ordinary particles through some nonrenormalizable interactions.
The most plausible candidate for the modulus is the dilaton in
string theories and it decays most likely into two photons:
$\phi \rightarrow 2\gamma $\cite{K-Y}
\footnote{
The modulus decay into two neutrinos is suppressed 
since it has a chirality flip and vanishes for massless neutrinos
\cite{K-Y}.
Similarly to the two photon decay the modulus may decay 
into two gluons.
For such a case the lifetime of the modulus may be shorter than that in eq.
(\ref{lifetime-moduli}) by a factor about 9.
}
through a coupling to two photons written as
\begin{eqnarray}
    \label{int-modui}
    {\cal L}_{int} ~=~
    \frac{ b }{ 4 M_G } \phi F_{\mu \nu} F^{\mu \nu}.
\end{eqnarray}
Here, we have introduced a dimension-less parameter $b$ which depends on
the type of superstring theories and compactifications%
\footnote{
For example, the dilaton has a coupling $b$ = $\sqrt{2}$\cite{dilaton}
for a compactification of the M-theory\cite{M-theory}.
}.
In the present analysis, we take $b$ as a free parameter of the order one
representing the various compactifications in string theories.
Then, the lifetime of the modulus is given by
\begin{eqnarray}
    \label{lifetime-moduli}
    \tau_\phi 
    ~\simeq~ \frac{ 64 \pi }{ b^2 } \frac{ M_G^2 }{ m_\phi^3 }
    ~\simeq~ 7.6 \times 10^{23} ~\mbox{sec} ~
       \frac{ 1 }{ b^2 } 
       \left( \frac{ 1 ~\mbox{MeV} }{ m_\phi } \right)^3.
\end{eqnarray}
Thus if $m_\phi \gg $ 100 MeV 
the modulus would decay within the age of the present universe
and its energy would be diluted 
below the critical density of the universe.
But for such a case another stringent constraint
should be considered.

\subsection{Constraint from the cosmic $X(\gamma)$-ray backgrounds}
Even if the lifetime of the modulus is longer than the age of the 
universe,
the modulus particle decays into photons in the past universe.
Thus, the produced radiation will contribute to the cosmic
$X(\gamma)$-ray backgrounds
and the observed backgrounds give a constraint 
on the mass and the lifetime of modulus $\phi$\cite{K-Y,H-K-Y}.

The photon number flux induced by the modulus decay is given by\cite{K-Y}
\begin{eqnarray}
  \label{photon-flux}
  F_\gamma(E_{\gamma}) ~&=&~ \frac{E_{\gamma}}{4\pi} \int_0^{t_0} dt ~
  \frac{2 n_\phi}{\tau_\phi} ( 1+ z ) 
  \delta \left( E_{\gamma} (1+z) - m_\phi/2 \right),
\\
  &\simeq&~
  \frac{ n_{\phi,0} }{ 2 \pi \tau_\phi H_0 }
  \left( \frac{ 2 E_{\gamma} }{ m_\phi } \right)^{3/2}
    \exp \left[ - \frac{ 2 }{ 3 \tau_\phi H_0 }
    \left( \frac{ 2 E_{\gamma} }{ m_\phi } \right)^{3/2}
  \right],
\end{eqnarray}
where $t_0$ is the age of the universe, $z$ the red-shift,
$H_0$ the present Hubble parameter and $E_{\gamma}$ the 
energy of $X(\gamma)$-ray.
And $n_{\phi,0}$ denotes the present number density 
of the modulus if it would be stable.
Here, we have assumed that the present total density parameter is
$\Omega_0 \simeq 1$. The detailed derivation of the photon flux
in eq.(\ref{photon-flux}) will be given in Appendix.

We may obtain a constraint on $\Omega_\phi h^2$ 
($\Omega_\phi = m_\phi n_{\phi,0}/\rho_c$) by requiring that
the maximum value of the flux in eq.(\ref{photon-flux}) 
should not exceed the observed $X(\gamma)$-ray backgrounds%
\cite{Xray1,Xray2,Xray3}.
The observational data are fitted by the following power-low 
spectra \cite{K-Y}
\begin{eqnarray}
    \label{obs-photon-flux}
    \frac{F_{\gamma, obs}(E_\gamma)}{\mbox{cm$^2\cdot$sr$\cdot$sec}} ~\simeq~ 
    \left\{
        \begin{array}[]{l r c l}
            8~(E_\gamma/\mbox{keV})^{-0.4} 
            & 0.1\mbox{keV} 
            & \lesssim E_\gamma \lesssim
            & 25\mbox{keV}\\
            380~(E_\gamma/\mbox{keV})^{-1.6} 
            & 25\mbox{keV} 
            & \lesssim  E_\gamma \lesssim 
            & 350\mbox{keV}\\
            2~(E_\gamma/\mbox{keV})^{-0.7} 
            & 350\mbox{keV} 
            & \lesssim  E_\gamma \lesssim 
            & 1\mbox{MeV}\\
            1.6\times 10^{-2} ~ (E_\gamma/\mbox{MeV})^{-1.8} 
            & 1\mbox{MeV} 
            & \lesssim  E_\gamma \lesssim 
            & 20\mbox{MeV}\\
            1.5\times 10^{-3} ~ (E_\gamma/\mbox{MeV})^{-1} ~~~~~~
            & 20\mbox{MeV} 
            & \lesssim  E_\gamma \lesssim 
            & 10\mbox{GeV}\\
        \end{array}
    \right. .  
\end{eqnarray}
The result is also shown in Fig.1.
We find that all mass region $10^{-2}$ keV $\lesssim m_{\phi} \lesssim$ 10 GeV
is excluded completely by the observed $X(\gamma)$-ray backgrounds.

In summary, we have shown that the thermal inflation could not dilute
the energy density of the modulus sufficiently so as to lower it
below the critical density 
of the present universe, if the modulus is stable.
Moreover, the observed cosmic $X(\gamma)$-ray backgrounds put 
a more stringent bound on the modulus with mass 
200 keV $\lesssim m_{\phi} \lesssim$ 10 GeV even if modulus is unstable.
Therefore, the cosmological string moduli problem {\it is not} solved
by the thermal inflation 
for the moduli masses less than ${\cal O}(10)$ GeV.
This excludes the mass range predicted in 
gauge-mediated SUSY-breaking theories
as long as $m_\phi\simeq m_{3/2}$.
%
\section{Thermal inflation without the R-axion decay and
         the cosmological moduli problem}
%
\subsection{Modified thermal inflation model}

If one sees carefully our scenario of the thermal history described in
the previous section, one might recognize that
the reason why the thermal inflation mechanism has failed to solve
the moduli problem could be attributed to the R-axions produced
by the flaton decay
\footnote{
In hidden sector models for the SUSY breaking the flaton decay
into R-axions may not be allowed energetically
(see footnote \ref{hidden-sector}).
Even if it is allowed the R-axions decay just after the flaton decay,
since the R-axion mass is of the order $m_\phi$, which reheats
the universe immediately. In any case sufficiently large entropy is
produced and the moduli problem may be solved\cite{Lyth-Stewart}.
}.
Indeed, the energy density of the relativistic R-axions
decreases faster than that of non-relativistic particles,
and as a consequence the R-axion decay into photons releases
much less entropy than in the case that the decay mode of
the flaton into R-axions is not open. Thus we might expect that
some mass regions in gauge-mediated SUSY breaking theories could survive
the cosmological constraints
if we could improve the model to forbid energetically
the flaton decay into R-axions.

Besides the stringent constraints discussed in the previous section,
there is another difficulty, i.e. the domain wall problem,
which we have ignored so far.
The origin of this problem is the degenerate minima of the potential
which possesses the discrete symmetry ${\bf Z}_{n+3}$.
Thus, in order to eliminate the domain walls we have to add
a small term to the potential which breaks the discrete symmetry explicitly.

One of economical modifications of the model which
satisfies both of the above two requirements is, for instance,
to add a linear term in the superpotential
\begin{eqnarray}
  \label{w-dm}
  \delta W ~=~ \alpha X,
\end{eqnarray}
which breaks the ${\bf Z}_{n+3}$ symmetry completely together with
the constant term $C$.
To collapse the domain walls before its energy dominates the universe,
the dimensionful parameter $\alpha$ is required 
to be\cite{Vilenkin}
\begin{eqnarray}
  |\alpha| \gtrsim \frac{ m_{3/2}^2 m_\chi M }{ M_{pl}^2 }.
  \label{const-dm}
\end{eqnarray}

The above explicit breaking term (\ref{w-dm}) in the superpotential
modifies the low energy potential of the flaton 
in eq.(\ref{flaton-potential}) as
\begin{eqnarray}
    \label{veff-dm}
    V_{eff}(X) &=& V_0 - 2 \frac{\alpha C}{M_G^2} ( X + X^\ast )
         - m_0^2 |X|^2 
         +  \frac{ \alpha }{ M_\ast^n } ( X^{n+2} + X^{\ast n+2 } )
\nonumber \\
  && + \frac{ n }{ n+3 } \frac{ C }{ M_G^2 M_\ast^n }
          ( X^{n+3} + X^{\ast n+3 } )
     + \frac{ 1 }{ M_\ast^{2n} } | X |^{ 2n +4 }.
\end{eqnarray}
In the following, instead of $\alpha$ we use a dimensionless
parameter $x$ defined by
$\alpha = - x M^{n+2}/M_\ast^n$, for simplifying the expressions below
\footnote{
We can write down the relation of $x$ and $\alpha$ 
in terms of the parameters appearing in the superpotential as
\begin{eqnarray}
    \alpha = - \frac{x}{[ (n+2) (1-x) ]^{ \frac{ n+2 }{ 2(n+1) } }} 
    m_0^{ \frac{ n+2 }{ n+1 } }
    M_\ast^{\frac{n}{n+1}},
\end{eqnarray}
up to the terms of ${\cal O}(m_{3/2}/m_0)$.
}.
Then the vev of the flaton is given by
\begin{eqnarray}
    \langle X \rangle \equiv M \simeq \left[ \frac{ 1 }{ (n+2)(1-x) } 
    \right]^{ \frac{ 1 }{ 2(n+1) } }
    ( m_0 M_\ast^n )^{\frac{ 1 }{ n+ 1}},
\end{eqnarray}
and $V_0$ is 
\begin{eqnarray}
    V_0 \simeq \frac{ n (1-x) + 1 }{ ( n+2 )( 1-x ) } m_0^2 M^2.
\end{eqnarray}

It should be noted that the explicit breaking term (\ref{w-dm}) 
does not affect much the dynamics of thermal inflation.
In the early universe the potential of the flaton 
near the origin takes the form as (see eq.(\ref{effective-potential}).)
\begin{eqnarray}
    V_{eff}(X) \simeq
    ( c T^2 - m_0^2 ) 
    \left| X + \frac{ 2 \alpha m_{3/2} }{ c T^2 - m_0^2 } \right|^2
    + V_0 - \frac{ 4 \alpha^2 m_{3/2}^2 }{ c T^2 - m_0^2 }
    + \cdots.
\end{eqnarray}
Although at the high temperature $T \gtrsim T_c \sim m_0$
the flaton does not sit at the origin,
the position of the minimum is
\begin{eqnarray}
    \langle X \rangle \simeq 
    - \frac{ 2 \alpha m_{3/2} }{ cT^2 - m_0^2 }
    \simeq \frac{ 2 x }{ \sqrt{ (n+2)(1-x) } } 
    \frac{ m_0 m_{3/2} }{ cT^2 - m_0^2 } ~ M,
\end{eqnarray}
and the deviation from the origin is 
suppressed by a factor of ${\cal O}(m_{3/2}/m_0)$
compared to the vev $M$ for the true minimum
\footnote{
If we take the value of $x$ to be very close to unity
($x < 1$ is always satisfied by the definition of $x$),
the position of the minimum at the high temperature is far from the origin.
Thus we consider the model with $x$ not so close to unity.
},
as far as $cT^2 - m_0^2 \gtrsim m_0^2 $.
Furthermore, the deviation of the vacuum energy of the flaton from $V_0$ is
estimated as
\begin{eqnarray}
    \delta V_0 \simeq - \frac{4 x^2 }{ n(1-x) + 1 }
    \frac{ m_{3/2}^2 }{ c T^2 - m_0^2 } V_0,
\end{eqnarray}
which has a suppression factor of ${\cal O}(m_{3/2}^2/m_0^2)$
and is consequently negligible.
Therefore, we find that if we take $x$ to be not so close to unity
the dynamics of the thermal inflation is not modified much
by adding the explicit breaking term (\ref{w-dm}).

Since the linear term (\ref{w-dm}) in the superpotential
breaks also an approximate $U(1)_R$ symmetry,
the term gives additional contributions not only to the flaton mass
but also to the $R$-axion mass.
Then their masses are given by
\begin{eqnarray}
    m_\chi^2 &\simeq& \frac{ 2(n+1) - n x }{ 1-x } m_0^2,
\\
    m_a^2 &\simeq&  \frac{ (n+2) x }{ 1-x } m_0^2.
\end{eqnarray}
Therefore, for the region%
\footnote{
Although this seems to be a highly restricted region,
one can see by taking the original parametrization by $\alpha$
that it actually corresponds to a broad one, namely
\begin{eqnarray}
    (-\infty) < \alpha < - \frac{ 2 (n+1) }{ 5n+8 }
    \left( \frac{ 5n+8 }{ 3 ( n+2)^2 } 
    \right)^{\frac{n+2}{ 2(n+1) } }
    M_\ast^{ \frac{n}{ n+1} } m_0^{ \frac{n+2}{n+1} }.\nonumber
\end{eqnarray}
}
\begin{eqnarray}
    \label{const-x}
    x_{min} \equiv \frac{ 2(n+1) }{ 5n+8 } < x ~( <1 ),
\end{eqnarray}
the flaton decay into two $R$-axions is kinematically
forbidden.
In the following we consider the thermal inflation model
with the explicit breaking term (\ref{w-dm})
satisfying eq.(\ref{const-x}),
and re-examine whether the model can solve the 
cosmological moduli problem in gauge-mediated 
SUSY breaking theories.

\subsection{Cosmological moduli problem with the modified thermal inflation}

First we argue the thermal history after the thermal inflation.
At the end of the thermal inflation ($T \sim T_c$)
the flaton field begins to oscillate around the true minimum
of the potential (\ref{veff-dm}),
and when the Hubble parameter becomes comparable to the total width
of the flaton, the flaton decays only into SUSY standard model particles
since the decay into $R$-axions is not allowed.
As discussed in section 2, 
the flaton decays dominantly into Higgs particles 
if kinematically allowed, 
or when the flaton mass is smaller than the threshold of 
the decay into two Higgs particles
the flaton decays into two photons.
Each width is represented in eqs.(\ref{chi-2higgs}) and (\ref{chi-2photon})
, respectively.
In both cases, the flaton energy is transferred to radiation
and reheats the universe immediately.
Then the modified thermal inflation model 
increases the entropy by a factor given by putting $\epsilon_a = 0$
and denoting $T_{R,SM}$ by $T_{R}$ in eq.(\ref{delta-sm}),
\begin{eqnarray}
    \label{delta-dm}
    \Delta \simeq 1 + \frac{ 4 }{ 3 } 
    \frac{ V_0 }{ ( 2 \pi^2/45) g_\ast T_c^3 T_R },
\end{eqnarray}
where the reheating temperature $T_R$ 
is obtained by the decay width of the 
flaton.

For the case that the flaton can decay into Higgs particles
($m_\chi > 130$ GeV)
\footnote{
From the experimental lower bound on the mass for Higgs bosons
\cite{LEP}, we take the Higgs mass
to be 65 GeV in order to obtain as conservative constraints as possible.
},
$T_R$ is represented as
\begin{eqnarray}
    \label{tr-higgs}
    T_R &\simeq& 0.14~
    F(C_h)
    \frac{ m_\chi^{3/2} M_G^{1/2}}{M},
\\
    &\simeq& 21 ~\mbox{GeV}~ 
    F(C_h)
    \left( \frac{ m_\chi }{ 100 ~\mbox{GeV} }\right)^{3/2}
    \left( \frac{ 10^{10} ~ \mbox{GeV} }{ M } \right),
\end{eqnarray}
where $F(C_h)$ is defined by
\begin{equation}
  \label{Function}
  F(C_h) \equiv
     \left[C_h+2\left(\frac{\alpha_{em}}{4\pi}\right)^2\right]^{1/2}.
\end{equation}
Thus in this case, as pointed out in Ref.\cite{Lyth-Stewart},
the reheating temperature $T_R$ can be taken to be high enough 
($T_R \gtrsim 10$ MeV) to maintain the success of big bang
neucleosynthesis.
On the other hand for the case $m_\chi \leq 130$ GeV,
the reheating temperature is estimated from eq.(\ref{chi-2photon}) as
\begin{eqnarray}
    \label{tr-gamma}
    T_R &\simeq&
    1.1 \times 10^{-4} ~\frac{ m_\chi^{3/2} M_G^{1/2} }{M},
\\
    &\simeq& 17 ~\mbox{ MeV} ~
    \left( \frac{ m_\chi }{ 100~\mbox{GeV} } \right)^{3/2}
    \left( \frac{ 10^{10}~\mbox{GeV} }{ M } \right).
\end{eqnarray}

Now let us estimate the cosmic energy density
of coherent modulus oscillation in the present universe.
The present energy density of the ``big-bang'' modulus
is given by replacing $\Delta$ in eq.(\ref{pre-rhobb}) by the expression 
of the entropy production (\ref{delta-dm}), hence
\begin{eqnarray}
    \left( \frac{ \rho_\phi }{ s } \right)_{BB}
    \simeq 3.8 ~
    \left( \frac{ T_c }{ m_\chi } \right)^3
    \left( \frac{ \phi_0 }{ M_G } \right)^2
    \frac{ m_\phi^{1/2} m_\chi^{3} M_G^{1/2} T_R }{ V_0 }.
\end{eqnarray}
On the other hand, the present energy density of the ``thermal-inflation''
modulus is estimated from eq.(\ref{pre-rhoti}) as
\begin{eqnarray}
     \left( \frac{ \rho_\phi }{ s } \right)_{TI}
    \simeq 0.38 ~
    \left( \frac{ \phi_0 }{ M_G } \right)^2
    \frac{ V_0 T_R }{ m_\phi^2 M_G^2 }.
\end{eqnarray}
Then we turn to estimating the lower bound of the total 
energy density of the modulus.
The lower bound is given by\cite{H-K-Y}
\begin{eqnarray}
    \left( \frac{ \rho_\phi }{ s } \right)_0
    \simeq \mbox{max} 
    \left[ \left( \frac{ \rho_\phi  }{ s } \right)_{BB},
        \left( \frac{ \rho_\phi  }{ s } \right)_{TI}
    \right]
    &\geq &
    \sqrt{ \left( \frac{ \rho_\phi  }{ s } \right)_{BB}
        \left( \frac{ \rho_\phi  }{ s } \right)_{TI} },
\nonumber \\
   \label{ab-130g}
    &\simeq& 1.2 \left( \frac{ \phi_0 }{ M_G } \right)^2
    \left( \frac{ T_c }{ m_\chi } \right)^{3/2}
    \frac{ m_\chi^{3/2} T_R }{ m_\phi^{3/4} M_G^{3/4} }.
\end{eqnarray}
Here, the equality is satisfied when
$(\rho_\phi/s)_{BB} = (\rho_\phi/s)_{TI}$, i.e. when
\begin{eqnarray}
    \label{bb=ti-130l}
    m_\chi \simeq \frac{ 2.5 \times 10^2 }{ C_{V_0}^{2/7} }
    \left( \frac{ T_c }{ m_\chi } \right)^{3/7}
    m_\phi^{5/14} M_G^{1/14} T_R^{4/7},
\end{eqnarray}
where $C_{V_0} = \frac{ n(1-x) + 1 }{ (n+2)[ 2(n+1) - n x]}$ and
we have used eq.(\ref{tr-gamma}) for an expression of $T_R$,
assuming $m_\chi \leq$ 130 GeV.
Then, the lower bound of the moduli energy is estimated from 
eq.(\ref{ab-130g}) as
\begin{eqnarray}
    \left( \frac{\rho_\phi}{s} \right)_0
    \gtrsim \frac{ 4.8 \times 10^3}{ C_{V_0}^{3/7} }
    \left( \frac{ \phi_0 }{ M_G } \right)^2
    \left( \frac{ T_c }{ m_\chi } \right)^{15/7}
    \frac{ T_R^{13/7} }{ m_\phi^{3/14} M_G^{9/14} }.
\end{eqnarray}
Thus the lowest reheating temperature $T_R \simeq 10$ MeV
and $x=x_{min}$
give the minimum abundance as
\begin{eqnarray}
    \label{min-130l1}
    \Omega_\phi h^2 \gtrsim 3.9 \times 10^{-4}
    \left[ \frac{ 8(n+2)^2 }{3n+8} \right]^{3/7}
    \left( \frac{ 1 ~\mbox{GeV} }{m_\phi} \right)^{3/14},
\end{eqnarray}
for $\phi_0 \simeq M_G$ and $T_c \simeq m_\chi$.
However, if we put $T_R \simeq 10$ MeV in eq.(\ref{bb=ti-130l})
we observe that the assumption $m_\chi < 130$ GeV holds only when 
\begin{eqnarray}
    \label{const-mp}
    m_\phi \lesssim m_c \equiv 5.2 \times 10^{-2} ~\mbox{GeV}~
    C_{V_0}^{4/5} 
    \left( \frac{ m_\chi }{ T_c } \right)^{6/5}.
\end{eqnarray}
Thus, for $m_\phi \lesssim m_c$ the theoretical lower bound
(\ref{min-130l1}) is realized in the case $m_\chi <$ 130 GeV
and the main decay mode is $\chi \rightarrow 2\gamma$.

On the other hand, for the modulus mass $m_\phi > m_c$ we can obtain
the minimum abundance lower than the r.h.s.~of eq.(\ref{min-130l1})
by making use of eq.(\ref{tr-higgs}) which is applicable for $m_\chi >$ 130GeV
instead of eq.(\ref{tr-gamma})
\footnote{
In this case $m_{3/2} > m_c$. Then, the flaton decay into Higssinos
should be forbidden, otherwise the gravitinos produced in
succesive decay of Higgsinos overclose the present universe.
}. 
Therefore, for $m_\phi > m_c$
the condition that the lower bound is saturated in eq.(\ref{ab-130g}) is
\begin{eqnarray}
    \label{bb=ti-130g}
    m_\chi \simeq
    \frac{ 4.3 }{ [C_{V_0} F(C_h)]^{2/7} }
    \left( \frac{ T_c }{ m_\chi } \right)^{3/7}
    m_\phi^{5/14} M_G^{1/14} T_R^{4/7}.
\end{eqnarray}
The lowest possible values
for the flaton mass and the reheating temperature
satisfying eq.(\ref{bb=ti-130g}) are practically given by
$m_\chi \simeq$ 130 GeV and $T_R$ = 10 MeV, yielding the minimum abundance
in eq.(\ref{ab-130g}) as
\begin{eqnarray}
    \left( \frac{\rho_\phi}{s} \right)_0
    \gtrsim 2.9 \times 10^{-13} ~
    \left( \frac{ \phi_0 }{ M_G } \right)^2 
    \left( \frac{ T_c }{ m_\chi } \right)^{3/2}
    \left( \frac{ 1~\mbox{GeV} }{ m_\phi } \right)^{3/4}.
\end{eqnarray}
Comparing with the present critical density in eq.(\ref{critical-density}),
$\phi_0 \simeq M_G$ and $T_c \simeq m_\chi$ leads to
\begin{eqnarray}
    \label{min-130g}
    \Omega_\phi h^2 \gtrsim 8.1 \times 10^{-5}
    \left( \frac{ 1~\mbox{GeV} }{ m_\phi } \right)^{3/4}.
\end{eqnarray}

In Fig.2 we show the lower bound for the enegy density of the modulus
predicted from eqs.(\ref{min-130l1}) and (\ref{min-130g}).
A remarkable consequence which distinguishes itself from the result in
section 3 is that for all the moduli mass region
$10^{-2}$ keV $\lesssim m_\phi \lesssim$ 10 GeV, the predicted lower bound
can be taken to be below the critical density of the present universe.

So far, we have been considering the case $m_\phi > H_{TI}$.
Then let us discuss briefly what happens in the case $m_\phi < H_{TI}$.
From eqs.(\ref{ab-2}) and (\ref{delta-dm}), 
the present abundance of such a modulus is given by
\begin{eqnarray}
  \label{ab-ati-dm}
  \left( \frac{ \rho_\phi }{ s } \right)_0
    ~\simeq~
    \frac{1}{8} \left( \frac{ \phi_0 }{ M_G } \right)^2
    T_R,
\end{eqnarray}
for any flaton mass.
Here we have assumed that the modulus mass is larger than
the decay width of the flaton (See the footnote \ref{assumption}.).
Comparing with the case $m_\phi > H_{TI}$,
the abundance in eq.(\ref{ab-ati-dm}) is always
greater than the minimum abundances in eqs.(\ref{min-130l1})
and (\ref{min-130g}) for $m_\phi \simeq 10^{-2}$ keV--1 GeV.

In Fig.2 we also show the constraint from the observed $X(\gamma)$-ray
backgrounds which is derived in section 3.
Then we can see that it excludes the modulus mass region
500 keV $\lesssim m_\phi \lesssim$ 10 GeV.
Therefore, we conclude that the cosmological problems are resolved
in a class of gauge-mediated SUSY-breaking models
with the gravitino mass $10^{-2}$ keV $\lesssim m_{3/2} \lesssim$ 500 keV,
provided that the modified thermal inflation takes place
and the flaton decay into R-axions is forbidden.
We should comment here that the ansatz in eq.(\ref{eq-c}) is
always satisfied for the parameter region we have analyzed.

%
\section{Discussion and conclusions}
%
We have found in the first part of our analysis that due to
the presence of too light R-axions
the original thermal inflation model\cite{Lyth-Stewart} could not resolve
infamous cosmological moduli problems in
gauge-mediated SUSY-breaking theories,
and all the gravitino mass region, 
$10^{-2}$ keV $\lesssim m_{3/2} \lesssim$ 10 GeV, is excluded
as long as $m_\phi \simeq m_{3/2}$ is fulfilled.

In the next step, we have modified the original thermal inflation model
so that the R-axion decay of flaton is kinematically suppressed,
and as a result we have succeeded to make a region
$10^{-2}$ keV $\lesssim m_{3/2} \lesssim$ 500 keV to survive
the cosmological constraints. Moreover, it can be shown that
a small window 1 GeV $\lesssim m_{3/2} \lesssim$ 10 GeV appears,
if we take $\phi_0/M_G \simeq $ 0.01--0.1 that is nevertheless
a slightly improbable condition
\footnote{
This window will also appear if one adopts very small values for
$b \simeq $ 0.01--0.1 in eq.(\ref{int-modui})
}.
One should note, however, that this window is located in the region
$m_{3/2} \gtrsim {\cal O}(1)$ MeV, in which it is far from easy
to construct mechanisms to produce a sufficient number of baryons
consistent with today's observational data\cite{G-M-M,K-Y}
if one adopts the original Affleck-Dine mechanism\cite{Affleck-Dine}.
Here, it may be interesting that an enough number of baryons will be created
even for $m_{3/2} \gtrsim {\cal O}(1)$ MeV if we adopt a variant type of
Affleck-Dine baryogenesis discussed in Ref.\cite{S-K-Y},
since gauge-mediated models with $m_{3/2} \sim $ 1 GeV are
relatively easily constructed\cite{H-I-Y,P-T}

In the present analysis we have adopted the effective potential
(\ref{flaton-potential}) for the flaton. 
We consider that the soft mass $m_0^2$ in eq.(\ref{flaton-potential})
is induced by the SUSY-breaking effects and hence
the potential (\ref{flaton-potential}) is applicable
only for the scale below masses $m_\psi$ of messenger multiplets.
Therefore, $\langle X \rangle = M$ should be less than
$m_\psi (\simeq 10^{14 - 15} m_{3/2}$ \cite{Dine-Nelson-Shirman}).
We have checked that this constraint is indeed satisfied for
the mass region $10^{-2}$ keV $\lesssim m_{3/2} \lesssim$ 10 GeV
we have studied.
(Note that $\langle X \rangle \simeq$ $10^7$ GeV -- $10^{12}$ GeV
for $m_{3/2} \simeq 10^{-2}$ keV -- 10 GeV.)

We would like to emphasize that the moduli could be the dark matter
in the present universe with their masses being
possibly in the region $10^{-2}$ keV $\lesssim m_\phi \lesssim$ 100 keV.
Fig.3 exhibits the region of the parameters
$m_\chi$ and $M_*$ in which the conditions $\rho_\phi \leq \rho_c$
and $T_R \gtrsim$ 10 MeV are satisfied for $m_\phi =$ 100 keV.
By noting that this region contains
plausible values $m_\chi \sim 10^1$--$10^2$ GeV and $M_* \sim 10^{18}$ GeV
fortunately, one would be encouraged to expect that the above emphasis
is by no means a nonsensical suggestion. The detailed study on
the spectrum of the $X(\gamma)$-ray emitted from the cosmic moduli
will be given in Ref.\cite{A-H-K-Y}.

As concerns the moduli masses, it is another intriguing possibility that
so-called ``MeV-bump'', an excess in the $\gamma$-ray background spectrum
around $\sim$ 1 MeV\cite{Xray2}, might be a signal for
the moduli decay\cite{K-Y}. Anyway, future analyses of
$X(\gamma)$-ray backgrounds in these mass regions are expected
to provide us with more detailed information about the moduli\cite{Kamae}.

If the existence of light moduli fields is a generic prediction of
string theories, it will probably present us with a plenty of subjects in
low energy physics, typical examples of which are nothing but
the cosmological moduli problems we have attempted to solve in this
paper.

\begin{center}
{\bf ACKNOWLEDGEMENT}
\end{center}

We would like to thank T. Kamae, H. Murayama and S. Orito
for the encouragement.

\appendix

\section{Photon flux}

Since the velocity dispersion of the moduli is negligible, two
monochromatic photons with energy $m_{\phi}/2$ are produced in the
moduli decay. 
Thus, the spectrum $S(E')$ of the photon per decay is written as
\begin{equation}
        S(E') = 2\delta(E' - m_{\phi}/2).
\end{equation}
The number density $n_{\phi}(z)$ of the moduli at red-shift $z$ is given by
\begin{equation}
        n_{\phi}(z) = n_{\phi,0}(1+z)^{3}\exp(-t/\tau_{\phi}).
\end{equation}
Then the present flux from the moduli is estimated as
\begin{eqnarray}
    F_{\gamma}(E_{\gamma})  & = &
    \frac{E_{\gamma}}{4\pi} \int_{0}^{t_0} dt'
    \frac{1}{\tau_{\phi}} n_{\phi}(z) (1+z)^{-3}
    \frac{dE'}{dE_{\gamma}}S(E') \nonumber \\
    & =  & 
    \frac{E_{\gamma}}{4\pi} \int_{0}^{t_0} dt'
    \frac{1}{\tau_{\phi}} n_{\phi,0}\exp(-t'/\tau_{\phi})
    (1+z) 2 \delta (E_{\gamma}(1+z) -m_{\phi}/2),
     \label{flux}
\end{eqnarray}
where $t_0$ is the present time, $E_{\gamma}$ the present energy 
of the photon and the factor $(1+z)^{-3}$ in the 
first equation represents the dilution due to the cosmic expansion.  Here 
we have taken into account that the photon with energy $E_{\gamma}$ had 
energy $E' = (1+z)E_{\gamma}$ when it was produced at the decay time. 
The red-shift $z$ is related to the cosmic time $t$ by 
\begin{equation}
    dt/dz  = -H_0^{-1} (1+z)^{-5/2} 
    [ \Omega_0 + (1-\Omega_0-\Omega_{\Lambda})/(1+z)
     + \Omega_{\Lambda}/(1+z)^{3}]^{-1/2},
\end{equation}
where $H_0$ is the present Hubble parameter, $\Omega_0$ the present
(total) density parameter and $\Omega_{\Lambda}$ the density
parameter of the cosmological constant. Then photon flux is
given by
\begin{eqnarray}
    F_{\gamma}(E_{\gamma}) & = & \frac{ n_{\phi,0} }{ 2 \pi \tau_\phi H_0 }
    \left( \frac{ 2 E_{\gamma} }{ m_\phi } \right)^{3/2}
    f(m_{\phi}/2E_{\gamma})\nonumber \\
    & \times & \exp \left[ - \int^{m_{\phi}/2E_{\gamma}}_0 d(1+z)
      \frac{1}{H_0\tau_{\phi}} (1+z)^{-5/2} f(1+z)\right],
    \label{appen-flux}
\end{eqnarray}
where 
\begin{equation}
     f(x) =    [ \Omega_0 + (1-\Omega_0-\Omega_{\Lambda})/x
     + \Omega_{\Lambda}/x^{3}]^{-1/2}.
\end{equation}
For $\Omega_0=1$ and $\Omega_{\Lambda}=0$, eq.(\ref{appen-flux}) is
simplified as
\begin{equation}
     F_{\gamma}(E_{\gamma})  = \frac{ n_{\phi,0} }{ 2 \pi \tau_\phi H_0 }
    \left( \frac{ 2 E_{\gamma} }{ m_\phi } \right)^{3/2}
    \exp \left[ -\frac{2}{3H_0\tau_{\phi}} 
    \left(\frac{ 2 E_{\gamma} }{ m_\phi } \right)^{3/2}\right].
\end{equation}
The flux $F_{\gamma}$ takes the  maximum value
$F_{\gamma,max}$ at
\begin{equation}
    E_{max}   = \left\{\begin{array}{ll}
          \frac{m_{\phi}}{2}~~~~~~~ & {\rm for}~~\tau_{\phi} > 
          \frac{2}{3}H_{0}^{-1} \\
          \frac{m_{\phi}}{2} \left(\frac{3\tau_{\phi}H_{0}}{2}\right)^{2/3} 
          ~~~~~~~& {\rm for}~~\tau_{\phi} < \frac{2}{3}H_{0}^{-1}
          \end{array}\right. .
\end{equation}


\begin{figure}
    \centerline{\psfig{figure=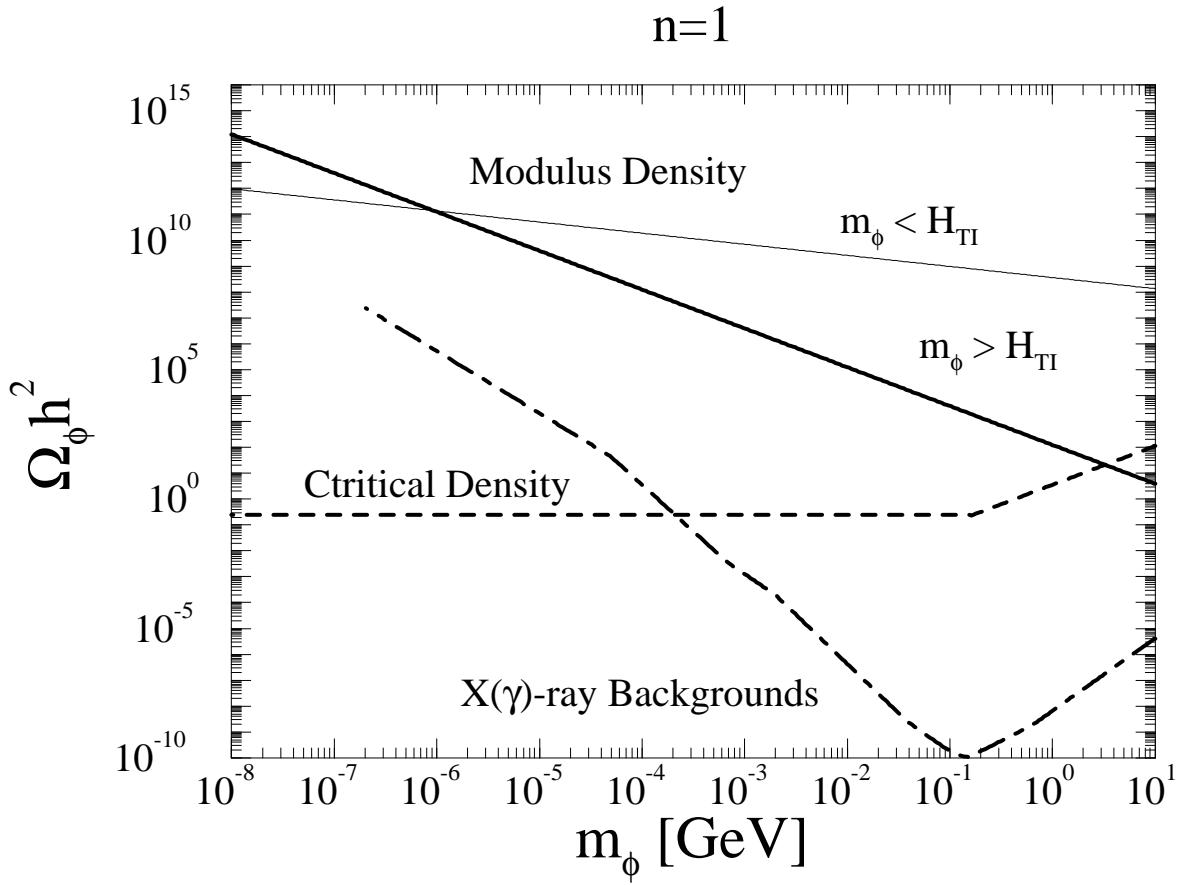,width=19cm}}
    \vspace{0.5cm}
\caption{
The lower bounds on the modulus densities $\Omega_\phi h^2$ 
in the original thermal inflation model for $n=1$.
The thick (thin) solid line represents the lower bound for the modulus density
with $m_\phi > H_{TI}$ ($m_\phi < H_{TI}$).
The upper bound from the present critical density
is represented by the dashed lines with a kink at
$m_\phi \sim$ 100 MeV. The kink appears because the constraint becomes
weaker for $m_\phi \protect\gtrsim$ 100 MeV
due to the modulus decay. The experimental upper bound
from the cosmic $X(\gamma)$-ray backgrounds ($b=1$)
is shown by dot-dashed line.}
\end{figure}

\begin{figure}
    \centerline{\psfig{figure=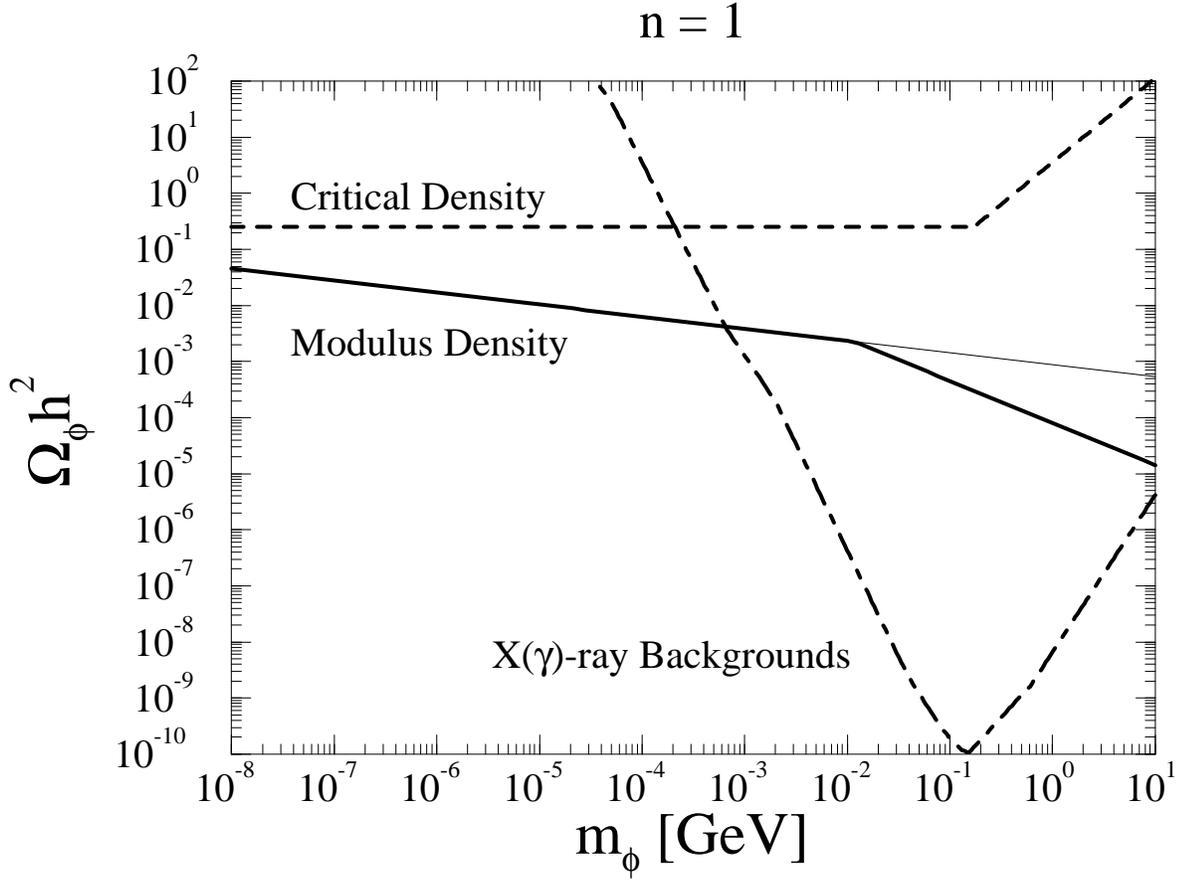,width=19cm}}
    \vspace{0.5cm}
\caption{
The lower bounds on the modulus densities $\Omega_\phi h^2$ 
in the modified thermal inflation model for $n=1$
where the flaton to two R-axions decay is forbidden.
The thick solid line with a kink at $m_\phi \sim$ 10 MeV
represents the lower bound for the modulus abundance.
The thin solid line for $m_\phi \protect\gtrsim$ 10 MeV
represents the lower bound for the case 
the flaton to Higgs fields decay is forbidden.
The upper bound from the present critical density
is represented by the dashed line.
The experimental upper bound  from the cosmic $X(\gamma)$-ray
backgrounds ($b=1$) is shown by dot-dashed line.}
\end{figure}

\begin{figure}
  \centerline{\psfig{figure=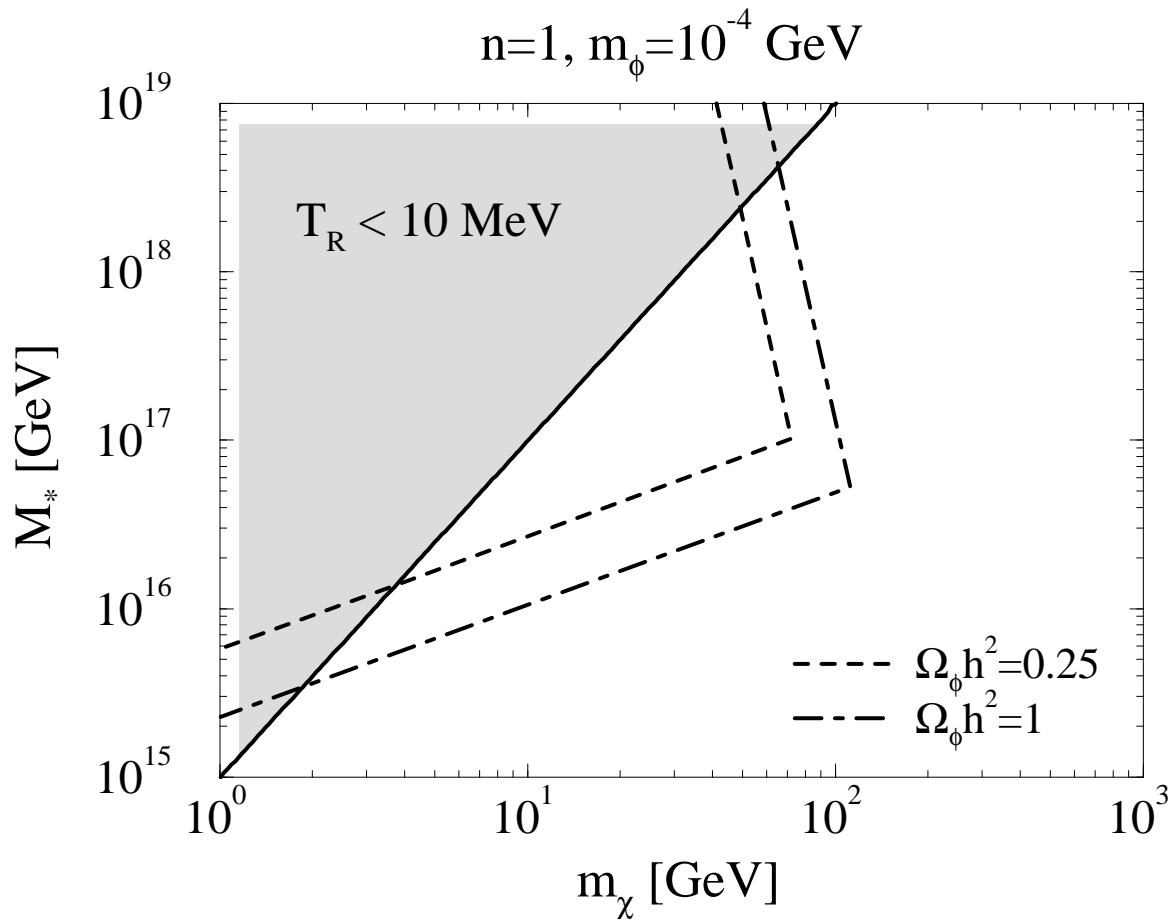,width=19cm}}
  \vspace{0.5cm}
\caption{
The contours of $\Omega_\phi h^2$ 
in the modified thermal inflation model for $n=1$
where the flaton to two R-axions decay is forbidden.
We take the modulus mass $m_\phi = 10^{-4}$ GeV.
The dashed line denotes $\Omega_\phi h^2$ = 0.25, and
the dot-dashed line denotes $\Omega_\phi h^2$ = 1.
In the shadow region $T_R <$ 10 MeV in which the big bang
nucleosynthesis is not operating well.
}
\end{figure}

\end{document}